\documentclass[twocolumn]{jpsj2} %% two-column layout

\title{%
Symmetry of Photoexcited States and Large-Shift Raman Scattering in Two-Dimensional Mott Insulators
}

\author{%
T. Tohyama\thanks{E-mail address: tohyama@imr.tohoku.ac.jp}
}

\inst{%
Institute for Materials Research, Tohoku University, Sendai, 980-8577
}

\recdate{August 22, 2005}

\abst{%
Symmetry of photoexcited states with two photoinduced carriers in two-dimensional Mott insulators is examined by applying the numerically exact diagonalization method to finite-size clusters of a half-filled Hubbard model in the strong-coupling limit.  The symmetry of minimum-energy bound state is found to be $s$-wave, which is different from a $d_{x^2-y^2}$ wave of a two-hole pair in doped Mott insulators.  We demonstrate that the difference is originated from an exchange of fermions due to the motion of a doubly occupied site.  Correspondingly large-shift Raman scattering across the Mott gap exhibits a minimum-energy excitation in the A$_1$ ($s$-wave) channel.  We discuss implications of the results for the Raman scattering and other optical experiments.
}

\kword{%
Photoexcited state, Hubbard model, Mott insulator, Raman scattering, Exact diagonalization
}

\begin{document}
\maketitle

\section{Introduction}

The charge gap in Mott insulators is a consequence of strong electron correlation represented by large on-site Coulomb interaction.  The correlation induces novel phenomena in terms of the interplay of charge and spin degrees of freedom.~\cite{Maekawa}  Photoexcitation across the Mott gap induces two carriers, an unoccupied site and a doubly occupied site of electrons.  In two dimensions (2D), the motion of the two carriers is strongly affected by the presence of localized spins in the background: The propagation of a carrier is known to induce a spin cloud around the carrier as a consequence of the misaligned spins along the carrier-hopping paths, and the two carriers prefer to form bound states by minimizing the loss of magnetic energy due to the spin cloud.~\cite{Tohyama}  The formation of such bound states is similar to the case of two holes introduced into 2D Mott insulators. The bound state formed in the two-hole ground state has $d_{x^2-y^2}$ symmetry (equivalently the B$_1$ representation in the $D_4$ group), provided that the exchange interaction between localized spins is not small compared with the nearest-neighbor hopping amplitude.~\cite{Chernyshev}  This is considered to be related to $d_{x^2-y^2}$ superconducting symmetry in the high-$T_c$ cuprates.

The nature of photoexcited states in 2D Mott insulating cuprates such as Sr$_2$CuO$_2$Cl$_2$ has been examined by using linear~\cite{Choi,Lovenich} and nonlinear optical response experiments~\cite{Schumacher,Ashida,Kishida,Maeda} and large-shift Raman scattering experiments across the Mott gap.~\cite{Liu,Salamon}  Among these experimental techniques, the large-shift Raman scattering is the best one to see the symmetry of the photoexcited states directly.  However, not only excitations related to the bound states in the photoexcited states but also $d$-$d$ transitions between $3d$ orbitals contributes to the Raman scattering.~\cite{Salamon}  Therefore, it is important to give information on the contribution of the bound states to the large-shift Raman scattering.

There are several theoretical studies related to the photoexcited states and their symmetry in 2D Mott insulators.~\cite{Khveshchenko,Wrobel,M.Takahashi,A.Takahashi1,A.Takahashi2} It has been shown analytically~\cite{Wrobel} and numerically~\cite{M.Takahashi} that the bound states with odd-parity (dipole-allowed states) are higher in energy than those with even-parity (dipole-forbidden states).  However, symmetry of the lowest-energy bound state has not been clarified yet, although two possibilities, the A$_2$~\cite{Khveshchenko} or B$_1,$~\cite{Wrobel} have been proposed. Therefore, it is important to clarify the symmetry of the bound sate in order to fully understand the nature of the photoexcited states.  Also it is interesting whether the symmetry is related to that of a bound state of two-hole pair in  doped Mott insulators.

In this paper, we theoretically examine symmetry of photoexcited states and large-shift Raman scattering in the 2D Mott insulators.  We apply a numerically exact diagonalization method to finite-size clusters of a half-filled Hubbard model in the strong-coupling limit.  In the calculations, we introduce various boundary conditions. An averaging procedure over twisted boundary conditions is also used to reduce finite-size effects.  The symmetry of the lowest-energy bound state is found to be neither A$_2$ nor B$_1$, but A$_1$ ($s$-wave symmetry). We demonstrate that, if sign changes due to an fermion exchange caused by the motion of a doubly occupied site are not taken into account, the symmetry becomes B$_1$.  The large-shift Raman scattering exhibits a lowest-energy excitation in the A$_1$ channel.  This is different from the experimental data showing a lowest-energy excitation with A$_2$ symmetry~\cite{Liu,Salamon} and thus supports a proposal that the A$_2$ excitation is due to a $d$-$d$ transition from $d_{x^2-y^2}$ to $d_{xy}$ orbitals.~\cite{Salamon}

The rest of this paper is organized as follows.  We introduce an effective Hamiltonian of the half-filled 2D Hubbard model in the strong-coupling limit, and show outlines of the procedure for choosing boundary conditions in \S~\ref{Model}.  In \S~\ref{Symmetry}, calculated results of the distribution of photoexcited states are shown and the symmetry of the lowest-energy bound state is discussed.  In \S~\ref{RamanScattering}, we show calculated spectra of the large-shift Raman scattering and compare them with experiments.  The summary is given in \S~\ref{Summary}.

\section{Model and Method}
\label{Model}

The Hubbard model is given by
\begin{equation}
H_\mathrm{Hub}=H_t+H_U
\label{Hub}
\end{equation}
with
\begin{equation}
H_t=-t \sum_{i,\delta,\sigma}\left( c^{\dagger}_{i,\sigma} c_{i+\delta,\sigma} + \mathrm{H.c.} \right)
\label{Hubt}
\end{equation}
and
\begin{equation}
H_U=U \sum_{i} n_{i, \uparrow} n_{i, \downarrow}\;,
\label{HubU}
\end{equation}
where $c^{\dagger}_{i,\sigma}$ is the creation operator of an electron with spin $\sigma$ at site $i$, $ n_{i, \sigma}=c^{\dagger}_{i,\sigma}c_{i,\sigma}$, the summation of $\delta$ runs over a pair of the nearest-neighbor sites between $i$ and $i+\delta$, $t$ is the nearest-neighbor hopping integral, and $U$ is the on-site Coulomb interaction.  In the strong coupling limit $(U \gg t)$, there is no doubly occupied site in the half-filled ground state, resulting in the Heisenberg model:
\begin{equation}
H_0=J \sum_{i,\delta} \left( \boldsymbol{S}_{i} \cdot \boldsymbol{S}_{i+\delta} - \frac{n_{i} n_{i+\delta}}{4} \right)\;, 
\label{Heisenberg}
\end{equation}
where $\boldsymbol{S}_{i}$ is the spin operator with $S=1/2$
 at site $i$, $n_{i}=n_{i, \uparrow} + n_{i, \downarrow}$, and $J=4t^2/U$.

Photoexcited states across the Mott gap have both one doubly occupied site and one vacant site. An effective Hamiltonian describing the photoexcited states is obtained by restricting the Hilbert spaces to a subspace with one doubly occupied site. By performing the second order perturbation with respect to the hopping term $H_t$, eq.~(\ref{Hubt}), the effective Hamiltonian is given by~\cite{M.Takahashi,A.Takahashi1}
\begin{equation}
H_\mathrm{eff}=\Pi_{1} H_t \Pi_{1}
- \frac{1}{U} \Pi_{1} H_t \Pi_{2} H_t \Pi_{1}
+ \frac{1}{U} \Pi_{1} H_t \Pi_{0} H_t \Pi_{1}
+U
\label{effectiveH}
\end{equation}
where $\Pi_{0}$, $\Pi_{1}$, and $\Pi_{2}$ are projection operators onto the Hilbert space with zero, one, and two doubly occupied sites, respectively.  We note that the effective model, eq.~(\ref{effectiveH}), can reproduce well the optical conductivity of the Hubbard model with large $U$ under periodic boundary conditions for the $N=18$ and 20 clusters~\cite{Tohyama3} and under antiperiodic boundary conditions for the $N=16$ cluster.\cite{Nakano}  Hereafter $\hbar=e=c=1$, $e$ and $c$ being the elementary charge and the speed of light, respectively, and the distance between the nearest-neighbor sites in the two-dimensional lattice is set to be unity.  Throughout this paper, we take $U/t=10$.

We use the exact diagonalization method based on the Lanczos algorithm to calculate low-lying eigenstate of the Heisenberg and effective Hamiltonians. In 2D systems, one uses a $N$-site square lattice with the translational vectors $\mathbf{R}_a=l\mathbf{x}+m\mathbf{y}$ and $\mathbf{R}_b=-m\mathbf{x}+l\mathbf{y}$, being that $N=l^2+m^2$ with integers $l,m\ge 0$.  Here, $\mathbf{x}$ and $\mathbf{y}$ are the vectors connecting nearest-neighbor sites in the $x$ and $y$ directions, respectively.  In this study, we take $N=18$ $(l=3, m=3)$, $N=20$ $(l=4, m=2)$, and $N=26$ $(l=5, m=1)$.

We impose periodic boundary conditions for the clusters along both the $x$ and $y$ directions. In addition, antiperiodic boundary conditions and mixed boundary conditions (one direction is periodic and the other is antiperiodic) are used to check the effect of the boundary condition on physical quantities.  In such small-size clusters, we are not free from finite-size effects that sometimes make the results unreliable.  In order to reduce the finite-size effects, we also introduce various boundary conditions with twist and average physical quantities over the twisted boundary conditions.   This procedure has been applied for various quantities in the $t$-$J$~\cite{Poilblanc} and $t$-$t'$-$t''$-$J$~\cite{Tohyama2} models.

The twist induces the condition that $c_{i+R_a,\sigma}=e^{\mathrm{i}\phi_a}c_{i,\sigma}$ and $c_{i+R_b,\sigma}=e^{\mathrm{i}\phi_b}c_{i,\sigma}$ with arbitrary phases $\phi_a$ and $\phi_b$.  Note that $\phi_a=\phi_b=0$ ($\pi$) corresponds to the periodic (antiperiodic) boundary conditions.
The phase $\phi_{a(b)}$ is defined as $\phi_{a(b)}=\boldsymbol{\kappa}\cdot\mathbf{R}_{a(b)}$
with an arbitrary momentum $\boldsymbol{\kappa}=\kappa_x\mathbf{x}+\kappa_y\mathbf{y}$.
$\boldsymbol{\kappa}$ usually scans an area surrounded by a square with four corners at $(\kappa_x,\kappa_y)=\pm\frac{\pi}{N}(l-m,l+m)$ and $\pm\frac{\pi}{N}(l+m,-l+m)$.  In order to perform the averaging procedure, we choose many $\boldsymbol{\kappa}$ in the square with equal intervals of $\pi/45$, $\pi/40$, and $\sqrt{2}\pi/8\sqrt{13}$ for the $N=$18, 20, and 26 clusters, respectively.  The total number of $\boldsymbol{\kappa}$, $N_\kappa$, results in $N_\kappa=$450, 320, and 64 for  $N=$18, 20, and 26, respectively.  The reason that $N_\kappa$ decreases with increasing $N$ is that time and memory for computing increase with $N$.  In particular, for $N=26$ we need to use supercomputing systems and the computing time of the optical conductivity for a $\boldsymbol{\kappa}$ point is about 130 minutes by using Hitachi SR11000 at the Institute for Solid State Physics, the University of Tokyo.  Therefore, it is time-consuming to increase $N_\kappa$ for $N=26$.
We note that imposing the twist is equivalent to transforming the operator $c^\dagger_{i,\sigma}c_{i+\delta,\sigma}$ into $\exp(\mathrm{i}\boldsymbol{\kappa}\cdot\boldsymbol{\delta})c^\dagger_{i,\sigma}c_{i+\delta,\sigma}$, $\boldsymbol{\delta}$ being the displacement vector from site $i$ to $i+\delta$. Therefore, the Heisenberg model, eq.~(\ref{Heisenberg}), is independent of the choice of $\boldsymbol{\kappa}$.

\section{Symmetry of Photoexcited States}
\label{Symmetry}

The regular part of the optical conductivity detects optical-allowed photoexcited states. The real part of the conductivity for a given $\boldsymbol{\kappa}$ is expressed as 
\begin{equation}
\sigma_\kappa(\omega)= \frac{\pi}{N\omega} \sum_m \left|\left< \Psi^{\boldsymbol{\kappa}}_m \left| j_x^{\boldsymbol{\kappa}} \right|0\right>\right|^2 \delta (\omega- E^{\boldsymbol{\kappa}}_m + E_0)\;,
\label{Regular}
\end{equation}
where $\left| 0\right\rangle$ is the ground state of the Heisenberg model with energy $E_0$, and $\left| \Psi^{\boldsymbol{\kappa}}_m\right\rangle$ represents a photoexcited state with energy $E^{\boldsymbol{\kappa}}_m$.  $j_x^{\boldsymbol{\kappa}}$ is the $x$ component of the current operator under the twist up to second order of $t$: 
\begin{eqnarray}
j_x^{\boldsymbol{\kappa}}&=&\mathrm{i}t\sum_{i,\delta,\sigma} \delta_{x} e^{-\mathrm{i}\boldsymbol{\kappa}\cdot\boldsymbol{\delta}} \tilde{c}^{\dagger}_{i+\delta,\sigma}
\tilde{c}_{i,\sigma}
+\mathrm{i}\frac{t^2}{U}\sum_{i,\delta,\delta^{\prime},\sigma}
\left( \delta_{x}-\delta^{\prime}_{x} \right) \nonumber \\
&&\times \Bigl( e^{\mathrm{i}\boldsymbol{\kappa}\cdot(\boldsymbol{\delta}^\prime-\boldsymbol{\delta})} \tilde{c}^{\dagger}_{i+\delta,\sigma} \tilde{c}^{\dagger}_{i,-\sigma} \tilde{c}_{i,-\sigma} \tilde{c}_{i+\delta^\prime,\sigma} \nonumber \\
&&\ \ - e^{\mathrm{i}\boldsymbol{\kappa}\cdot\boldsymbol{\delta}^\prime} \tilde{c}^{\dagger}_{i+\delta,\sigma} \tilde{c}^{\dagger}_{i,-\sigma}
\tilde{c}_{i+\delta,-\sigma} \tilde{c}_{i+\delta^\prime,\sigma} \Bigr)\;.
\label{current}
\end{eqnarray}
The creation and annihilation operators, $\tilde{c}^{\dagger}_{i,\sigma}$ and $\tilde{c}_{i,\sigma}$, are projected onto the subspace with either zero or one doubly occupied site. The $\delta_{x}$ is the $x$ components of $\boldsymbol{\delta}$.  For the periodic boundary conditions, $\boldsymbol{\kappa}$ is chosen to satisfy $\boldsymbol{\kappa}\cdot\mathbf{R}_a=\boldsymbol{\kappa}\cdot\mathbf{R}_b=0$ ($\pi$).  For the mixed boundary conditions, we take an average of $\sigma_\kappa(\omega)$ with $\boldsymbol{\kappa}\cdot\mathbf{R}_a=0$ and $\boldsymbol{\kappa}\cdot\mathbf{R}_b=\pi$ and that with  $\boldsymbol{\kappa}\cdot\mathbf{R}_a=\pi$ and $\boldsymbol{\kappa}\cdot\mathbf{R}_b=0$.  The optical conductivity under the averaging procedure is given by
\begin{equation}
\sigma_\mathrm{ave}(\omega)=\frac{1}{N_\kappa}\sum_{\boldsymbol{\kappa}} \sigma_\kappa(\omega)\;.
\label{RegularAve}
\end{equation}

\begin{figure}
\begin{center}
\includegraphics[width=8.5cm]{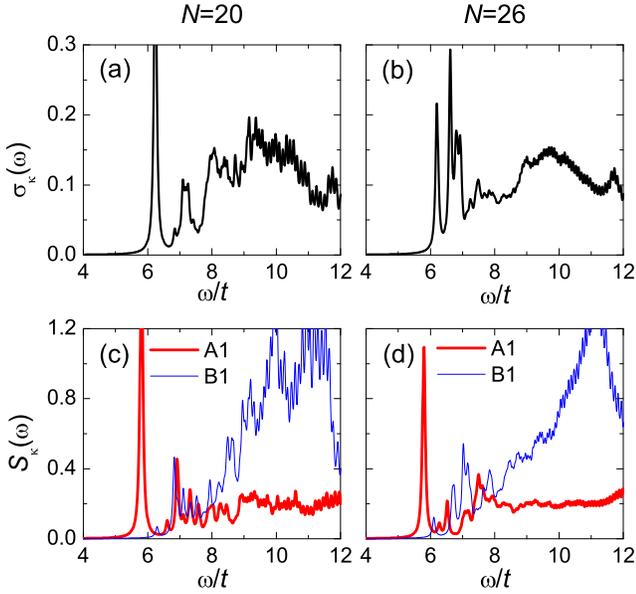}
%\vskip4cm
\caption{\label{fig1}
(Color online) Optical conductivity $\sigma_\kappa(\omega)$ for half-filled Hubbard clusters in the strong coupling limit with the size of $N=20$ (a) and $N=26$ (b), and dynamical correlation function $S_\kappa(\omega)$ of A$_1$-symmetry operator (the thick line) and of B$_1$-symmetry operator (the thin line) for $N=20$ (c) and $N=26$ (d).  $U/t=10$.  Periodic boundary conditions are employed and delta-functions are broadened by a Lorentzian with a width of $0.05t$.}
\end{center}
\end{figure}

\begin{figure}
\begin{center}
\includegraphics[width=8.5cm]{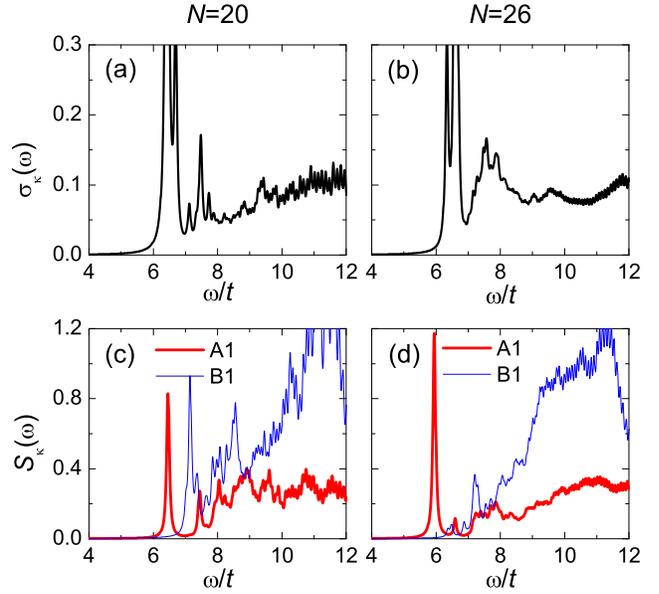}
%\vskip4cm
\caption{\label{fig2}
(Color online) The same as Fig.~\ref{fig1}, but under antiperiodic boundary conditions.}
\end{center}
\end{figure}

\begin{figure}
\begin{center}
\includegraphics[width=8.5cm]{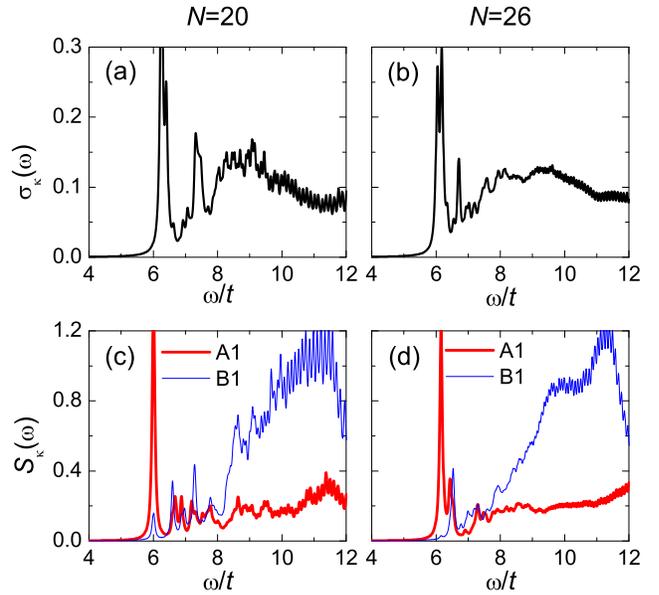}
%\vskip4cm
\caption{\label{fig3}
(Color online) The same as Fig.~\ref{fig1}, but under mixed boundary conditions, where the average of two cases [periodic (antiperiodic) along the $x$ ($y$) direction and antiperiodic (periodic) along the $x$ ($y$) direction] is taken.}
\end{center}
\end{figure}

\begin{figure}
\begin{center}
\includegraphics[width=8.5cm]{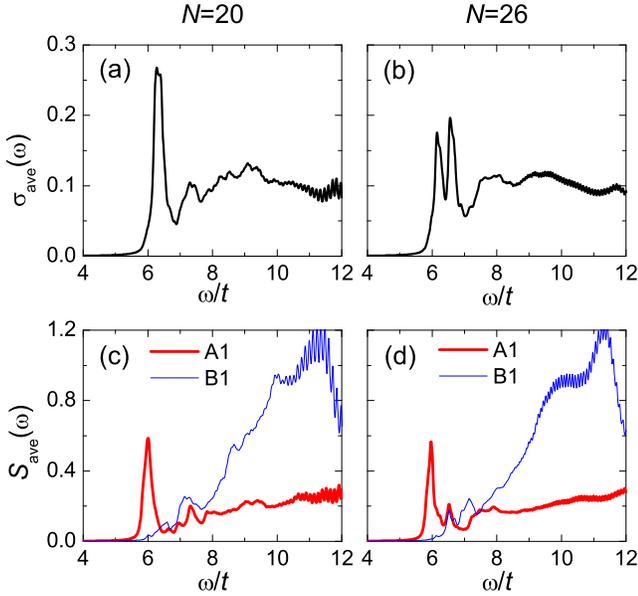}
%\vskip4cm
\caption{\label{fig4}
(Color online) The same as Fig.~\ref{fig1}, but an averaging procedure for twisted boundary conditions is employed.}
\end{center}
\end{figure}

We show the optical conductivity under various boundary conditions for the $N=20$ and $N=26$ clusters in the panels (a) and (b), respectively, of Figs.~\ref{fig1} - \ref{fig4}: Results for the periodic, antiperiodic, mixed, and averaged boundary conditions are shown in Fig.~\ref{fig1}, Fig.~\ref{fig2}, Fig.~\ref{fig3}, Fig.~\ref{fig4}, respectively.  Both the $N=20$ and $N=26$ cases in Figs.~\ref{fig1}(a) and \ref{fig1}(b) show a globally similar distribution of the spectral weight, exhibiting two prominent structures: One is a broad-peak structure centered at around $\omega=10t$, and the other is an absorption-edge structure at around $\omega=6t$ separated from the broad peak.  The latter structure is sensitive to the exchange interaction $J$,~\cite{Tohyama} and thus originates from magnetically induced bound states.  This global distribution of the spectral weight does not change under the antiperiodic and mixed boundary conditions as shown in Figs.~\ref{fig2} and \ref{fig3}, though fine structures in the spectra depend on the boundary conditions.

The similarity of the spectral-weight distribution between $N=20$ and $N=26$ is also seen in Figs.~\ref{fig4}(a) and \ref{fig4}(b), where we take the averaging procedure, eq.~(\ref{RegularAve}). The averaging is found to reduce a difference of the edge structure for $N=20$ and $N=26$ under the periodic boundary conditions~\cite{Onodera} [see Figs.~\ref{fig1}(a) and \ref{fig1}(b)]:  A single peak  appears at $\omega=6.2t$ for $N=20$, while for $N=26$ there are four peaks with similar magnitude within the region of $6t \lesssim \omega \lesssim 7t$.  After taking the averaging over $\boldsymbol{\kappa}$, the edge structure becomes wider for $N=20$ but narrower for $N=26$.  Therefore, the difference appearing in the cases of the periodic boundary conditions is reduced in Fig.~\ref{fig4}, although a two-peak structure still remains for $N=26$. The reduction demonstrates efficiency of the averaging procedure.  We note that $\sigma_\mathrm{ave}(\omega)$ of $N=18$ (not shown) is similar to that of $N=20$.

In order to examine whether optical-forbidden states are lower or higher in energy than allowed ones, we introduce operators with A$_1$ and B$_1$ symmetry composed of the nearest-neighbor hoppings:
\begin{equation}
C_\pm^{\boldsymbol{\kappa}}=\sum_{i,\sigma} \left( e^{-\mathrm{i}\kappa_x} \tilde{c}^\dagger_{i+x,\sigma}\tilde{c}_{i,\sigma} \pm e^{-\mathrm{i}\kappa_y}\tilde{c}^\dagger_{i+y,\sigma}\tilde{c}_{i,\sigma}+\mathrm{h.c.} \right)\;,
\label{C+-}
\end{equation}
where the plus ($+$) and minus ($-$) signs correspond to A$_1$ and B$_1$, respectively, and $\kappa_\alpha$ is the $\alpha$ component of $\boldsymbol{\kappa}$.  The dynamical correlation function of the operators for a given $\boldsymbol{\kappa}$ is given by
\begin{equation}
S_\kappa(\omega)=\frac{1}{N}\sum_m \left|\left< \Psi^{\boldsymbol{\kappa}}_m \left| C_\pm^{\boldsymbol{\kappa}} \right|0\right>\right|^2 \delta (\omega- E^{\boldsymbol{\kappa}}_m + E_0)\;,
\label{Sym}
\end{equation}
and an averaged correlation is expressed as
\begin{equation}
S_\mathrm{ave}(\omega)=\frac{1}{N_\kappa}\sum_{\boldsymbol{\kappa}} S_\kappa(\omega)\;.
\label{SymAve}
\end{equation}
We note that the final states $\left| \Psi^{\boldsymbol{\kappa}}_m \right>$ cannot be divided into subspaces of these representations because of the twisted boundary conditions.

$S_\kappa(\omega)$ for $N=20$ and $N=26$ is shown in the panels (c) and (d), respectively, of Figs.~\ref{fig1} - \ref{fig3}, and $S_\mathrm{ave}(\omega)$ is shown in Figs.~\ref{fig4}(c) and \ref{fig4}(d).  For the cases of the periodic, antiperiodic, and mixed boundary conditions (Figs.~\ref{fig1} - \ref{fig3}), we find edge-structures in the A$_1$ correlation separated from broad spectrum, indicating the presence of bound states.  We also find that in these boundary conditions the energy of the edge position of the A$_1$ correlation is lower than those of the B$_1$ correlation and $\sigma_\kappa(\omega)$.  As is the case of $\sigma_\mathrm{ave}(\omega)$, the averaging procedure makes differences between $N=20$ and $N=26$ smaller.  After averaging, the fact that the edge of A$_1$ is lower than that of B$_1$ is still preserved, indicating an intrinsic nature of the photoexcited states of the half-filled Hubbard model with large $U$.  As will be shown below, the states having A$_2$ and B$_2$ components are higher in energy than the A$_1$ dominated states.  These results indicate that the lowest-energy bound state in the photoexcited states of the Hubbard model has the A$_1$ symmetry, i.e., $s$-wave symmetry.

\begin{figure}
\begin{center}
\includegraphics[width=8.5cm]{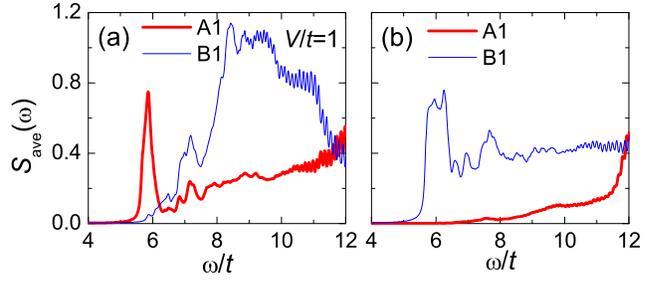}
%\vskip4cm
\caption{\label{fig5}
(Color online) Dynamical correlation function $S_\mathrm{ave}(\omega)$ of A$_1$-symmetry operator (the thick line) and of B$_1$-symmetry operator (the thin line) for a $N=20$ half-filled Hubbard cluster in the strong coupling limit. (a) $U/t=10$ and $V/t=1$.  (b) $U/t=10$, but sign changes due to the exchange of fermions caused by the motion of a doubly occupied site are neglected. An averaging procedure for twisted boundary conditions is employed and delta-functions are broadened by a Lorentzian with a width of $0.05t$.}
\end{center}
\end{figure}

This conclusion is not altered even if we include an attractive interaction, $V$, between neighboring doubly occupied and vacant sites, which induces an  excitonic bound state.  The resulting effective Hamiltonian is given by 
\begin{eqnarray}
H_V&=&H_\mathrm{eff}-V\sum_{i,\delta} \bigl[ n_{i,\uparrow}n_{i,\downarrow} \left(1-n_{i+\delta,\uparrow}\right)\left(1-n_{i+\delta,\downarrow}\right) \nonumber \\
&&\ \ \ \ \ + \left(1-n_{i,\uparrow}\right)\left(1-n_{i,\downarrow}\right) n_{i+\delta,\uparrow}n_{i+\delta,\downarrow}  \bigr]\;.
\label{effectiveHV}
\end{eqnarray}
Figure~\ref{fig5}(a) exhibits $S_\mathrm{ave}(\omega)$ for the $N=20$ cluster with $V=t$. We find no qualitative change of the bound states as compared with Fig.~\ref{fig4}(c): The excitation energy of the A$_1$ bound state is lower than that of B$_1$. 

In the two-hole doped Mott insulator, the hole pair forms a $d_{x^2-y^2}$ wave in the ground state.~\cite{White}  This is in contrast with the present results that the two-carrier pair produced by photoexcitation forms an $s$ wave.  It is important to clarify what is the origin of this difference.  We can easily notice a remarkably difference between the two-hole and two-photoexcited-carrier systems.  That is the difference of the electric charge of carriers.  In the photoexcited states, one of the carriers is not a hole but contains two electrons.  Therefore, in a basis representation where electrons are sorted in order of site, the motion of this carrier inevitably induces an exchange of fermions and gives an extra sign.  For instance, in the case of a two-site cluster with a doubly occupied site and a singly occupied site, the hopping of the doubly occupied site induces an extra sign:
$(-t c_{2,\uparrow}^\dagger c_{1,\uparrow}) c_{2,\downarrow}^\dagger c_{1,\downarrow}^\dagger c_{1,\uparrow}^\dagger \left| \mathrm{vac}\right> = +t c_{2,\downarrow}^\dagger c_{2,\uparrow}^\dagger c_{1,\downarrow}^\dagger \left| \mathrm{vac}\right>$, $\left| \mathrm{vac}\right>$ being the vacuum state.

In order to check the effect of the fermion exchange, we introduce an effective Hamiltonian that is the same as eq.~(\ref{effectiveH}) but reversing the sign of hopping of the doubly occupied site.  Figure~\ref{fig5}(b) shows $S_\mathrm{ave}(\omega)$ for the effective model. We find that the B$_1$ correlation exhibits bound states whose energies are lower than those of A$_1$, indicating the lowest-energy state to be $d_{x^2-y^2}$-wave.  From this result, we can conclude that the fermion-exchange process for the doubly occupied site plays a crucial role in making the A$_1$ bound state lowest in energy. 

\section{Large-Shift Raman Scattering}
\label{RamanScattering}

Large-shift Raman scattering for the half-filled Hubbard model is a good quantity to study the photoexcited states in the symmetry-resolved form. The Raman intensity for a given $\boldsymbol{\kappa}$ is given by
\begin{equation}
R_\kappa(\omega)=\frac{1}{ N} \sum_f \left|\left< \Psi^{\boldsymbol{\kappa}}_f \left| M_\mathrm{R} \right|0\right>\right|^2 \delta (\omega- E^{\boldsymbol{\kappa}}_f + E_0)
\label{Raman}
\end{equation}
with symmetry-resolved Raman operators $M_\mathrm{R}$: $M_\mathrm{A1}=M_{xx}+M_{yy}$, $M_\mathrm{A2}=M_{xy}-M_{yx}$, $M_\mathrm{B1}=M_{xx}-M_{yy}$, and $M_\mathrm{B2}=M_{xy}+M_{yx}$.  The operator $M_{\alpha\beta}$ ($\alpha,\beta=x,y$) is expressed as
\begin{equation}
M_{\alpha\beta}=\sum_m \frac{J_\beta^{\boldsymbol{\kappa}} \left|\Psi^{\boldsymbol{\kappa}}_m \right>\left<\Psi^{\boldsymbol{\kappa}}_m \right| j_\alpha^{\boldsymbol{\kappa}}}{E^{\boldsymbol{\kappa}}_m -E_0-\omega_\mathrm{i}+\mathrm{i}\eta}\;,
\label{RamanOpe}
\end{equation}
where $J_\beta^{\boldsymbol{\kappa}}$ connects two subspaces in the photoexcited states and is given by
$J_\beta^{\boldsymbol{\kappa}}=j_\beta^{\boldsymbol{\kappa}}+\mathrm{i} [U^{-1}\Pi_{1} H_t \Pi_{0} H_t \Pi_{1},\hat{\beta}]$,
where $\hat{\beta}$ is the $\beta$ component of the total position operator.  The Raman intensity under the averaging procedure is given by
\begin{equation}
R_\mathrm{ave}(\omega)=\frac{1}{N_\kappa}\sum_{\boldsymbol{\kappa}} R_\kappa(\omega)\;.
\label{RamanAve}
\end{equation}
We set the incident photon energy $\omega_\mathrm{i}$ to be $\omega_\mathrm{i}=U+6t=16t$ and focus on the Raman shift $\omega$ of order of $U$. We take the relaxation energy $\eta$ to be $0.2t$.

\begin{figure}
\begin{center}
\includegraphics[width=8.5cm]{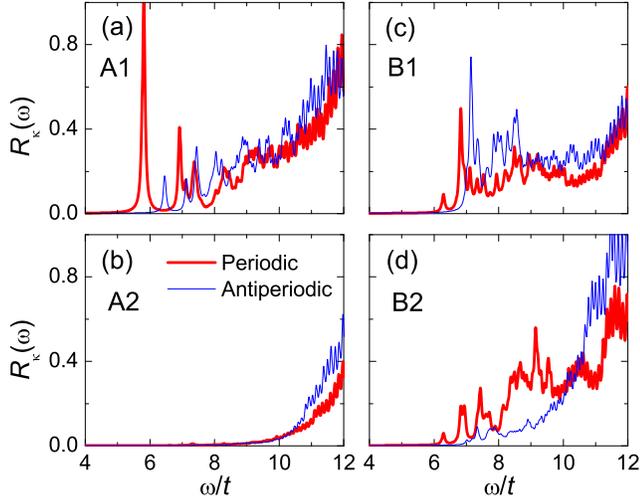}
%\vskip4cm
\caption{\label{fig6}
(Color online) Large-shift Raman scattering $R_\kappa(\omega)$ for a $N=20$ half-filled Hubbard cluster in the strong coupling limit. $U/t=10$. (a) A$_1$, (b) A$_2$, (c) B$_1$, and (d) B$_2$ scattering. The thick and thin lines represent the results under periodic and antiperiodic boundary conditions, respectively. Delta-functions are broadened by a Lorentzian with a width of $0.05t$.}
\end{center}
\end{figure}

\begin{figure}
\begin{center}
\includegraphics[width=8.5cm]{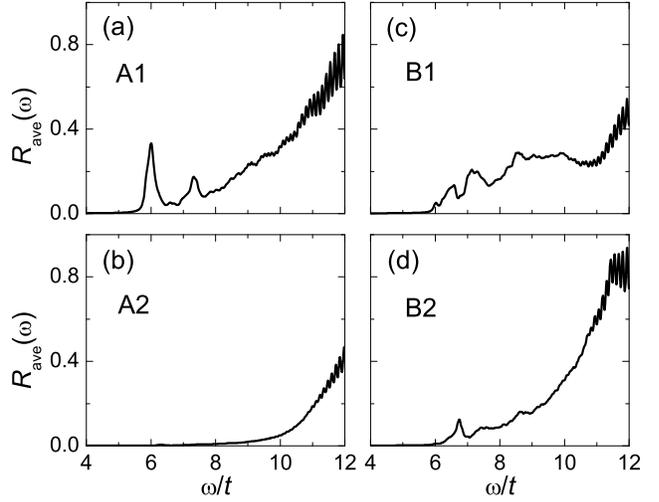}
%\vskip4cm
\caption{\label{fig7}
The same as Fig.~\ref{fig6} but an averaging procedure for twisted boundary conditions is employed.}
\end{center}
\end{figure}

Figure~\ref{fig6} shows $R_\kappa(\omega)$ for the $N=20$ cluster with the periodic and antiperiodic boundary conditions.  Although global features up to the high-energy region are similar between the two boundary conditions, edge structures of the spectrum depend on the boundary conditions.  In order to reduce the dependence, we use the averaging procedure, eq.~(\ref{RamanAve}).  $R_\mathrm{ave}(\omega)$ is shown in Fig.~\ref{fig7}.  Among all possible symmetry, the A$_1$ Raman scattering exhibits the lowest-energy excitation, being consistent with the A$_1$ correlation function discussed above.  For the B$_1$ and B$_2$ Raman scattering, spectral weights appear slightly above the A$_1$ scattering.  In the A$_2$ scattering, no weight is observed at the edge region around $\omega=6t$.  These results demonstrate again that the A$_1$ state is the lowest-energy state in the photoexcited states of the 2D Mott insulator. 

\begin{figure}
\begin{center}
\includegraphics[width=8.5cm]{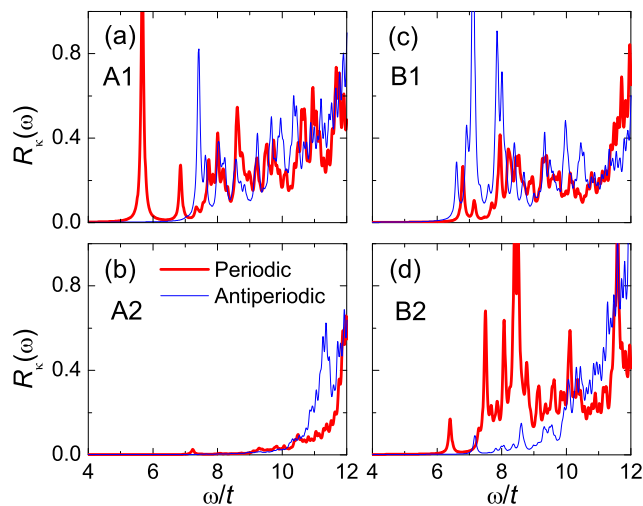}
%\vskip4cm
\caption{\label{fig8}
(Color online) The same as Fig.~\ref{fig6} but for $N=18$.}
\end{center}
\end{figure}

\begin{figure}
\begin{center}
\includegraphics[width=8.5cm]{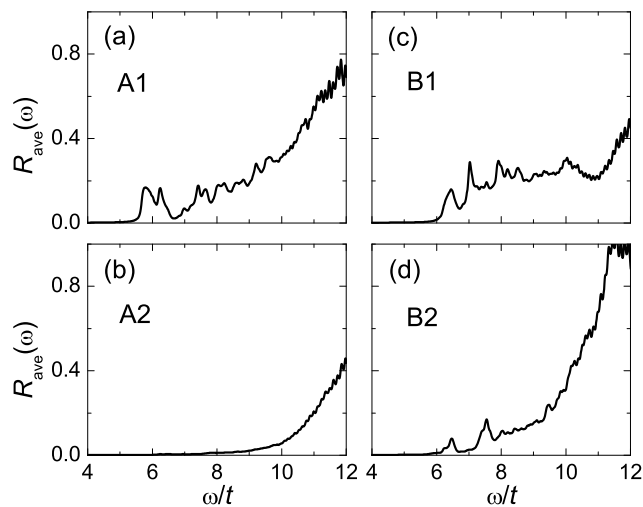}
%\vskip4cm
\caption{\label{fig9}
The same as Fig.~\ref{fig6} but for $N=18$ and an averaging procedure for twisted boundary conditions is employed.}
\end{center}
\end{figure}

Since the $N=20$ cluster is not a simple square lattice but a tilted one, it does not have the $D_4$ point group.  In order to check whether the data in Figs.~\ref{fig6} and \ref{fig7} are artifacts of tilted lattice or not, we examine $R_\kappa(\omega)$ and $R_\mathrm{ave}(\omega)$ for the $N=18$ cluster that has the $D_4$ group.  The results are shown in Figs.~\ref{fig8} and \ref{fig9}.  In the case of the periodic boundary conditions where the $D_4$ point group is fully held, $R_\kappa(\omega)$ exhibits the same behaviors as the case of $N=20$ under the periodic boundary conditions: The A$_1$ scattering shows the lowest-energy excitation and the A$_2$ scattering has no weight at around $\omega=6t$.  Although the difference of $R_\kappa(\omega)$ under the periodic and antiperiodic boundary conditions is larger in the case of $N=18$ than that in the case of $N=20$ shown in Fig.~\ref{fig6}, $R_\mathrm{ave}(\omega)$ obtained after the averaging in Fig.~\ref{fig9} shows spectral features similar to the case of $N=20$ in Fig.~\ref{fig7}.  This indicates that the behaviors of Raman intensities shown in Figs.~\ref{fig6} and \ref{fig7} are inherent in the square-lattice Hubbard model.

The large-shift Raman scattering experiments for parent compounds of high-$T_c$ superconductors such as Nd$_2$CuO$_4$ have shown that the A$_2$ scattering has the largest spectral weight whose excitation energy is lower than the absorption-peak position.~\cite{Salamon}  This is completely different from the present results shown in Figs.~\ref{fig7} and \ref{fig9}.  We consider that, as discussed in ref.~\citen{Salamon}, the observed large A$_2$ scattering would come from a $d$-$d$ transition from $d_{x^2-y^2}$ to $d_{xy}$ orbitals through photoexcited states.

\section{Summary}
\label{Summary}

Before summarizing our results, 
we discuss implications of the present results for recent two-photon absorption (TPA) experiment on Nd$_2$CuO$_4$.~\cite{Maeda}  In the experiment, there is a small absorption hump whose energy is about 0.3~eV lower than the optical (linear) absorption energy.  Combining this experimental data with third-harmonic generation data~\cite{Kishida,Kishida_priv} and analyzing the data by using a set of discrete excitation-energy levels, Kishida {\it et al.} have argued that the hump structure in TPA comes from a photoexcited state with B$_1$ symmetry and thus the lowest-energy photoexcitation is due to the B$_1$ symmetry.~\cite{Kishida_priv}  This is different from the present results that the minimum excitation comes from the A$_1$ symmetry.  According to the Kishida's analysis, it is difficult to assign the hump as the A$_1$ excitation.  Therefore, a possible interpretation is that the hump structure in TPA would originate from a $d$-$d$ excitation between, for instance, $d_{x^2-y^2}$ and $d_{3z^2-r^2}$ orbitals.  However, in order to come to final conclusions, we need further theoretical studies of, for instance, TPA spectra under the presence of various $d$ orbitals.

In summary, we have examined symmetry of photoexcited states and symmetry-resolved large-shift Raman scattering in the 2D Mott insulators by using an effective Hamiltonian of a half-filled Hubbard model in the strong-coupling limit and a numerically exact diagonalization method on finite-size clusters.  The symmetry of the lowest-energy bound state is found to be neither A$_2$ nor B$_1$, but A$_1$.  Therefore, the symmetry is different from that of a hole pair in doped Mott insulators.  We demonstrate that the difference is originated from fermion exchange induced by the motion of a doubly occupied site.  The large-shift Raman scattering exhibits a lowest-energy excitation in the A$_1$ channel.  This is different from the experiments showing a lowest-energy excitation with A$_2$ symmetry, which is probably due to a $d$-$d$ transition from $d_{x^2-y^2}$ to $d_{xy}$ orbitals. It seems to be necessary to include other $d$ orbitals into the photoexcited states for a complete understanding of the experimental data. This remains as a future problem.

\section*{Acknowledgment}

The author is very grateful to H. Kishida and H. Okamoto for their insightful comments on this work. Valuable discussions with K. Tsutsui and S. Maekawa are also appreciated.  This work was supported by a Grant-in-Aid for scientific Research from the Ministry of Education, Culture, Sports, Science and Technology of Japan, CREST, and NAREGI project.  The numerical calculations were partly performed in the supercomputing facilities in ISSP, University of Tokyo, and IMR, Tohoku University.

\end{document}